\documentclass[12pt]{iopart}
\usepackage{graphicx}
\begin{document}
\title[Report on the first binary black hole inspiral search in LIGO data] 
{Report on the first binary black hole inspiral search in LIGO data}

\author{Eirini Messaritaki\dag, for the LIGO Scientific Collaboration}

\address{\dag\ Center for Gravitation and Cosmology, 
University of Wisconsin-Milwaukee, Milwaukee, Wisconsin, 53201}


\begin{abstract}
The LIGO Scientific Collaboration is currently engaged in the first search for 
binary black hole inspiral signals in real data. We are using the data from 
the second LIGO science run and we focus on inspiral signals coming from binary
systems with component masses between 3 and 20 solar masses. We describe the 
analysis methods used and report on  preliminary estimates
for the sensitivities of the LIGO instruments during the second science run.
\end{abstract}





\section{Introduction}
\label{intro}

The Laser Interferometer Gravitational Wave Observatory (LIGO) 
consists of three Fabry-Perot-Michelson interferometers: a 4~km interferometer
at Livingston (hereafter L1), 
a 4~km interferometer at Hanford (hereafter H1) and a 2~km interferometer
also at Hanford (hereafter H2). Those instruments are
very close to the end of the commissioning phase and 
are approaching their design sensitivities as of December 
2004. The second science run (hereafter S2) of the LIGO intereferometers took
place from February 14 to April 14, 2003. 
During that run all three LIGO interferometers were operated in ``science 
mode'', 
stably and in coincidence. Even though the
design sensitivity goal had not been achieved at that time, the data
acquired represent the best broad-band sensitivity to gravitational
waves that had been achieved up to that date.

We report here on the search for gravitational waves from non-spinning 
binary black hole inspirals in the data from the second science run.
For that search we take advantage of the knowledge acquired via the binary
neutron star inspiral search and we use methods developed for it. However,
we have to address issues that are unique to binary black hole inspirals.

\section{Target sources}
\label{sources}

The target sources for the search described in this paper are binary
black hole systems with component masses equal to a few times the mass
of the sun. The intrinsic angular momentum of each component of
the binary is not taken into account. 
The effects of it will be examined in future searches.

Black hole binaries are believed to emit gravitational waves during 
all stages of their evolution, namely the inspiral, the merger 
and the ringdown. 
There are no reliable models for the waveform of the merger phase 
at present. Numerical work 
\cite{BakerCampanelliLoustoTakahashi,BakerCampanelliLousto}
has given insights into the merger problem and the progress in
numerical relativity indicates that models for the merger phase will
be developed in the future. 
At this time, it is more appropriate to search for binary black hole
mergers using techniques developed for finding unmodeled bursts
\cite{s1burst}. Additionally,
even though there are theoretical waveforms for the ringdown phase,
we do not search for ringdowns, for reasons that are explained below.
Thus, we focus on the gravitational waves emitted during the inspiral phase. 

At the first stages of the inspiral there is only small loss of energy per
orbit and the standard post-Newtonian
approximation for calculating the gravitational waves emitted 
is applicable. As the separation between the
two black holes becomes smaller, relativistic
effects become more significant and the standard post-Newtonian 
approach becomes less accurate. That specifically happens as the ``innermost
stable circular orbit'' (or ISCO) is approached, namely the last orbit 
before the highly non-adiabatic plunge of the black holes towards each 
other begins. The higher the masses of the two black holes, the earlier
the ISCO is reached. The test-mass approximation gives that 
the ISCO frequency can vary from 55 Hz for a system of two black holes with 
component masses equal to 40 $M_{\odot}$ up to 733 Hz for a system of two black
holes with component masses equal to 3 $M_{\odot}$.

Various different approaches based on post-Newtonian calculations 
\cite{LB,Standard,DIS,EOB}
have been followed in order to solve the difficult
problem of the late-time evolution of the inspiral phase of binary black
hole systems. 
These approaches differ in the approximation methods used to evolve the
system as it nears the ISCO and on the predicted ISCO frequency. 
Consequently, they also differ in
their predictions for the gravitational waveforms emitted by the
system at the last stages of the inspiral phase, when the frequency
is approaching the ISCO frequency.

A low-frequency cutoff of 100 Hz needed to be imposed on the S2 data,
for reasons that will be explained in sec.~(\ref{pipeline}). 
That cutoff limited the masses of the binary
systems for which we could detect an inspiral signal within the
LIGO band: we had to focus the search on binary black
hole systems with component masses between 3 and 
20 $M_{\odot}$. (The lowest ISCO frequency for that mass
range is that for the 20 - 20 $M_{\odot}$ system, equal to
110 Hz according to the test-mass approximation, 
just above the imposed low-frequency cutoff.)
For these binaries the expected ringdown frequencies range from 295 Hz to
1966 Hz \cite{ringdowns:Leaver}, mostly in the low-sensitivity
frequency band of LIGO during S2. That is the reason why we do not 
search for the ringdown phase of those binaries.

\section{Templates and filtering}
\label{templates}

The low-frequency cutoff imposed on the data does not only restrict
the mass range for this search but it also implies that for 
the black hole binaries of total mass close to 40 $M_{\odot}$
the LIGO interferometers were sensitive only 
to the very last part of the inspiral phase, during S2.
As was mentioned in sec.~(\ref{sources}), 
the various theoretical models produce different predictions and 
waveforms for that part of the evolution of the system.
Given that fact, we are faced with the dilemma of which one
of the physical waveforms to implement in the matched filtering
code that we use to search the data for inspiral signals. 
Since some of those waveforms differ significantly with each other (and
potentially with an \emph{actual} gravitational waveform),
choosing one that does not match well with the real inspiral gravitational wave
signal may hurt the efficiency of the search.

Alternatively, we choose to filter the data using a family of waveform 
templates which were developed specifically for detection of binary 
black hole inspirals.
Those templates were first described in a recent paper by
Buonanno, Chen and Vallisneri 
\cite{BCV} and are usually referred to as BCV templates.
The BCV templates are phenomenological templates because 
they are not derived by calculating the evolution of the binary system
according to a specific physical model. However, 
their phase evolution and amplitude are based on the post-Newtonian 
phase evolution and amplitude. They have
a high match with most physical waveform templates that have been
proposed in the literature and are known to be efficient at identifying
inspiral gravitational wave chirps in the data. Their frequency-domain form is
\begin{equation}
\tilde{h}(f) = f^{-7/6}\; (1 - \alpha f^{2/3} )\; \theta(f_{\mathrm{cut}} - f)
\; e^{i (2 \pi f t_0 + \phi_0 + \psi_0 f^{-5/3} + \psi_3 f^{-2/3})}.
\end{equation}
The amplitude component $f^{-7/6}$ is the standard restricted Newtonian
amplitude. The component $(\alpha \; f^{-1/2})$ is designed to capture
any post-Newtonian amplitude corrections and to obtain high matches with
non-adiabatic models that deviate from the Newtonian amplitude in the last
parts of the inspiral and the merger. The parameters $t_0$ and 
$\phi_0$ are the time of arrival and phase of the waveform. 
The parameters $\psi_0$ and $\psi_3$ are phenomenological parameters
governing the frequency evolution of the chirp, with values that are
related to the masses of the component black holes.
In order to obtain high matches with the various post-Newtonian models that 
predict different terminating frequencies, the
cutoff frequency $f_{\mathrm{cut}}$ is imposed to terminate the waveform.

We filter the Fourier-transformed data $\tilde{s}(f)$ with a template
$\tilde{h}(f)$ and construct the signal-to-noise ratio. In general
that is given by
\begin{equation}
\rho(h) = \frac{ <s,h> }{ \sqrt{<h,h>} }
\end{equation}
where the inner product is defined as
\begin{equation}
<s,h> = 4 \Re \int_0^{\infty} \frac{ \tilde{s}(f) \; \tilde{h}^{\ast}(f) }
{S_h(f)} df.
\label{snr}
\end{equation}
In eq.~(\ref{snr}), $S_h(f)$ is the one-sided noise power spectral density.
Specifically for the BCV templates, we construct a bank of templates for 
the intrinsic parameters $\psi_0$, $\psi_3$ and $f_{\mathrm{cut}}$
and we maximise the signal-to-noise ratio over the extrinsic
parameters $t_0$, $\phi_0$ and $\alpha$. 

A standard part of the matched filtering process is the
$\chi^2$-test \cite{chisqtest}. 
This test quantifies the degree to which the signal-to-noise
ratio is distributed in time in accordance to the way
it would be distributed for a true gravitational wave inspiral chirp.
It is very efficient at distinguishing binary
neutron star inspiral signals from loud non-gaussian noise bursts in the
data and is used in the binary neutron star inspiral search. However,
we found that test to not be applicable to the search for gravitational
waves from binary black hole inspirals in the S2 data. 
The expected short duration and  
small number of cycles in the S2 LIGO frequency band for many of the
possible
binary black hole inspiral signals made such a test unreliable unless
a very loose threshold were to be set. A loose threshold, on the other hand,
resulted in only a minimal reduction in the number of instrumental noise 
events picked up. That
and the fact that the $\chi^2$-test is computationally costly made us
decide to not use it in this search.

\section{Analysis pipeline and trigger generation}
\label{pipeline}

In order to search the data for binary black hole inspiral gravitational 
waves we have developed a multi-level pipeline. 
The pipeline is described in great detail in \cite{s2bns,s2bns:gwdaw}.
Technical 
issues such as segmentation of the data and power spectrum estimation 
are discussed in those papers and our choices for the
related parameters are explained. For completeness, a brief explanation 
of the main features of the pipeline is given here.

In order to increase the confidence that an event that comes out of
the pipeline is a real gravitational wave as opposed to an event
due to instrumental noise, we choose to search only S2 data 
acquired during times when the L1 interferometer and at least one of 
the two Hanford interferometers were operating in science mode. That
allows us to test the events that come out of the pipeline for coincidence
between the Hanford and Livingston sites. 
We choose to not search data from times when both H1 and H2 (but not L1) were
operating in science mode, due to the fact that noise bursts 
caused by environmental disturbances may be
correlated between H1 and H2 and may appear as 
coincident events at the end of our pipeline. We then have three
data sets to consider, based on which of the two Hanford interferometers
were operating: data from times when L1 and \emph{only} H1 were operating
(L1-H1 double coincident data), data from times when L1 and \emph{only} 
H2 were operating
(L1-H2 double coincident data) and data from times when L1 and \emph{both} H1 
and H2 were operating (L1-H1-H2 triple coincident data).

Various data quality cuts were imposed in order to make sure that
the data was appropriate for our analysis. Detailed investigations
resulted in us imposing the 100 Hz
low-frequency cutoff mentioned in sec.~(\ref{sources}). That cutoff
was necessary in order to eliminate the non-stationary noise that 
was present at frequencies around 70 Hz and was resulting in
a lot of spurious instrumental noise events being observed on the data.
We expect to lower this low-frequency cutoff in the analysis of future
science runs.

The data is broken up to ``analysis segments'' of 2048 s for power
spectrum estimation and for matched filtering.
One fundamental feature of the pipeline is the fact that we 
perform a triggered search. During the second science run, the
L1 interferometer was more sensitive than either of the Hanford
interferometers. More information on this will be given 
in sec.~(\ref{inject}).
Given that and the fact that the L1 interferometer is nearly aligned with the
Hanford interferometers, we expect that in most cases gravitational waves
observable in H1 and/or H2 are also observable in L1. For that reason,
we choose to start the triggered search by creating a template bank
over the parameters $\{ \psi_0, \; \psi_3, \; f_{\mathrm{cut}} \} $
for each 2048-s analysis segment of 
the L1 interferometer.
The number of templates in the bank varies with the variations in the
noise but is typically about 1000 templates.
The L1 data is filtered with that bank and
the times, phases, $\alpha$ and template parameters for 
which the signal-to-noise ratio exceeds the threshold
of 7 are recorded as ``triggers''. All
the corresponding triplets $\{ \psi_0, \; \psi_3, \; f_{\mathrm{cut}} \} $ 
form the ``triggered'' template bank with which the data from the
second interferometer (H1 for the L1-H1 and the L1-H1-H2 data or H2
for the L1-H2 data) is then filtered. The next step
is the coincidence step. We test if the triggers from
the two sites have consistent parameters and we reject those that do not. 
Specifically, we test that
the times of the triggers at the two sites differ by no more than the
travel time of gravitational waves between the two sites (10 ms) 
plus the accuracy by which we can recover the end time of the 
gravitational wave signals (10 ms), i.e. we require the trigger times
at the two sites
to be within $\pm$ 20 ms of each other. We also test
that the template parameters of the triggers are equal for the
time-coincident triggers that come from the two sites. For the cases
of double coincident data (L1-H1 or L1-H2) the coincident triggers
are recorded at this point. For the case of triple coincident data,
the triggers that survive L1-H1 coincidence are used to create
a second triggered bank with which the data from the H2 interferometer
is filtered. The H2 triggers are tested for coincidence
with the L1 and H1 triggers generated for the same triple coincident data.
However, given the limited sensitivity of the H2 interferometer, we 
keep all the L1-H1 coincident triggers regardless of whether they are 
coincident with any H2 triggers or not. We end up with either double
or triple coincident triggers for the triple coincident data. The
final step of the pipeline is the clustering of the triggers. Clustering
the triggers is necessary because both strong noise bursts and
strong gravitational wave signals are known to give large numbers of
triggers with different coalescence times and filter parameters, 
which are difficult to handle in the follow-up analysis. 
We choose to cluster the
triggers that occur within intervals of 250 ms, approximately half the
duration of
the longest possible binary black hole inspiral signal that we can
detect in the S2 data given the low-frequency cutoff of 100 Hz.

Finally, it should be noted that the various parameters of the pipeline
(signal-to-noise ratio threshold, coincidence time window,
etc.) were determined by adding simulated
inspiral signals in the data and trying to recover them with our 
pipeline. The tuning was done on a part of the S2 data (approximately 
10\%) that was predetermined and
representative of all the data from the second science run. That
subset of data is referred to as ``playground'' data.

\section{Monte-{C}arlo simulations}
\label{inject}

In order to tune the various parameters of the pipeline, 
to validate the pipeline and to verify that if there is a
detectable signal in the data we would be able to identify it, 
we perform Monte Carlo simulations during which we add simulated 
binary black hole inspiral signals into the data stream 
(in software) and try to recover them with our pipeline. 
Monte-Carlo simulations also allow us to determine the
efficiency of our pipeline at detecting binary black hole inspiral signals.
The added simulated signals are referred to as ``injections''.

\subsection{Characteristics of injected signals}
\label{charact}

As was mentioned in sec.~(\ref{templates}), the templates that we use for the
matched filtering step of the pipeline are phenomenological
templates which do not result from calculations based on physical models.
However, for the signals that we inject in the data, we 
use waveforms that \emph{are} based on physical models. Of the large
number of physical waveforms available in the literature, we 
choose to inject effective-one-body (EOB) \cite{EOB}, 
standard post-Newtonian (spN) \cite{Standard}
and PadeT1 waveforms (PT1) \cite{DIS}, all of second post-Newtonian order. 
In addition to validating and measuring the efficiency of our pipeline, the
fact that we inject different physical waveforms allows us to 
determine how effective the BCV templates are at identifying 
signals that are based on different models. 

In the case of neutron star binaries there are theoretical population
models that are based on radio observations of such systems. 
Those models restrict the presence of binary neutron stars into galaxies 
and give possible mass distributions for them. In contrast,
binary black hole systems can only be detected via the gravitational
waves that they generate and have never been observed before. Thus there are 
no observation-based theoretical models 
for their population. For that reason the binary black hole 
simulated signals
are not placed only inside galaxies but are spread throughout space.

The choice for the distances of the simulated
signals is based on the S2 sensitivities of the LIGO
instruments. Specifically to binary black hole inspirals,
theoretical calculations using the standard post-Newtonian waveforms
showed that given the \emph{actual} S2 power spectra for the three instruments,
the distance at which an optimally oriented 5 - 5 $M_{\odot}$
binary inspiral would be detected with a signal-to-noise ratio of
8 (effective range) varied from 4 to 7 Mpc for L1, 2 to 3 Mpc for
H1 and 1 to 2 Mpc for H2. In fact the effective range
increased for all three interferometers as the run was progressing.
Due to the large range of ISCO frequencies for the
black hole binaries we consider, these ranges can be quite different
for different signal-to-noise ratios and different masses of the systems.
Binaries of total mass close to 6 $M_{\odot}$ do not have enough power
in band
and will not be detectable at distances as large as the effective ranges
mentioned above. Binaries of total mass close to 40 $M_{\odot}$ do not
have many cycles in the S2 LIGO frequency band and will also not be
detectable at distances as large as the effective ranges mentioned above.
On the other had, binaries of total mass around 20 $M_{\odot}$ 
have enough power in band and can be
expected to be detectable at distances larger than the effective ranges
previously mentioned.

For our initial simulations we choose
to inject signals the distance of which varies between 100 kpc and 10 Mpc.
The random sky angles and orientations of the binaries
result in some signals having much larger effective distances.
Populations of distances as low as 10 kpc and as high as 20 Mpc 
will also be injected, but the results are not presented here.
Various choices for the distance distribution were considered. 
It was determined that
choosing a uniform-distance distribution or a uniform-volume 
distribution would overpopulate the
region of large distances (where the instruments were not very
sensitive) and would result in very few injections with small
and intermediate distances (for which we are more interested in
measuring the efficiency of our pipeline). For those reasons we choose 
a distribution that has uniform $\log$(distance).

We limit the component masses of the binaries for the simulated signals 
between 3 and 20 $M_{\odot}$. That choice
was justified in sec.~(\ref{sources}).
We inject populations with uniform component mass
between 3 and 20 $M_{\odot}$ as well as populations with uniform total mass
between 6 and 40 $M_{\odot}$ with each component mass between
3 and 20 $M_{\odot}$. The results presented here are for the latter choice.

\subsection{Analysis of {M}onte-{C}arlo simulations}

\begin{figure}[ht]
\begin{center}
\includegraphics[height=7cm,width=10cm,clip]{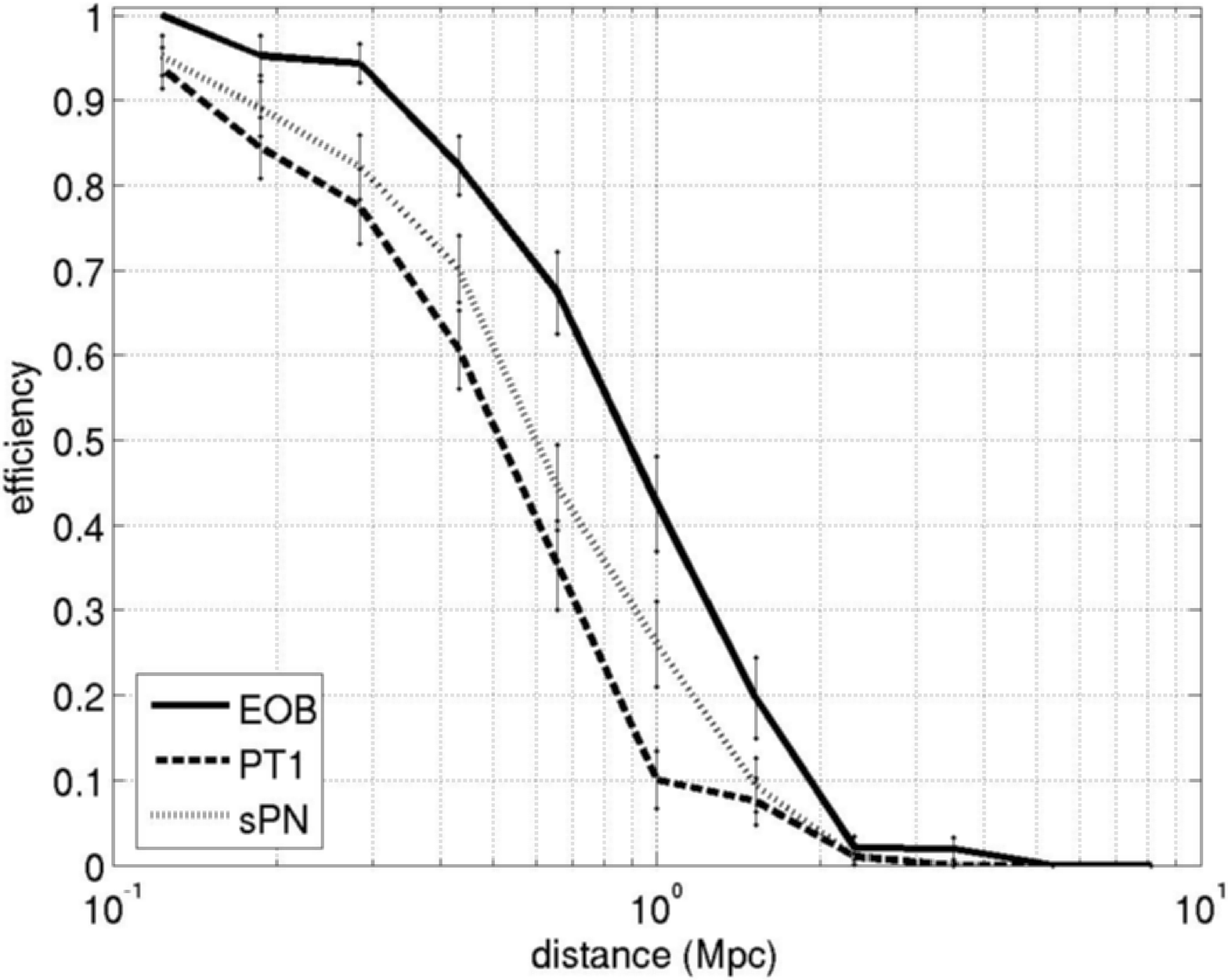}
\end{center}
\caption{\label{f:effic}
The efficiency versus the distance for the three
families of injected waveforms is shown. 
The binomial error bars are also shown.}
\end{figure}

We report here on injections performed on the playground part of
the S2 data. 
The efficiency for recovering the injected waveforms (number
of found injections of a given distance divided by the total
number of injections of that distance) is shown in fig.~(\ref{f:effic}).

The first interesting result is that we can detect gravitational waves
from binary black hole inspirals with efficiency of at least 10\% at 
distances of 1 Mpc, for the mass range that we are exploring. That is
consistent with the expected reach of the LIGO interferometers during S2.
The fact that some injected signals are missed at distances lower than 
the effective range mentioned in sec.~(\ref{charact})
can be attributed to various factors. Some of the injected 
signals have highly non-optimal orientation for at least one of
the LIGO interferometers and can only produce triggers with signal-to-noise
ratio lower than the threshold. Other signals, injected at times early
during S2, correspond to epochs during which the reach of the LIGO 
interferometers was limited. Finally, some signals correspond to 
very low or very high masses and thus 
are not expected to be detectable at distances as large as the
effective ranges mentioned previously.

The second interesting result pertains to the efficiency of our pipeline
for different injected waveforms. 
It is clear from fig.~(\ref{f:effic}) that the efficiency for
recovering EOB waveforms is higher than that for
standard post-Newtonian or PadeT1 waveforms, for all
distances.
The efficiency of recovering standard post-Newtonian waveforms is
a little higher than the efficiency for recovering PadeT1 waveforms.
That is consistent with the observation \cite{BCV} that the BCV
templates have higher overlap with the EOB waveforms than with the
standard post-Newtonian or the PadeT1 waveforms.

\section{Background estimation}
\label{background}

We estimate the rate of accidental coincidences 
(also known as background rate) for this search by introducing an artificial 
time offset (lag) $\Delta t$ to the triggers coming from the
Livingston detector relative to the Hanford detectors.  The time-lag
triggers are fed into the coincidence steps of the pipeline and,
for triple coincident data, to the step of the filtering of the H2 data
and the last coincidence.  For a given time lag, the
triggers which emerge from the end of the pipeline are considered as one
single trial representation of an output from a search if no
signals are present in the data.  By choosing a lag of more than
20~ms (equal to the time coincidence window between the two sites), 
we ensure that a true gravitational wave could never produce coincident
triggers in the time-shifted data streams. To avoid correlations,  we use
lags longer than the duration of the longest waveform ($\sim$ 0.6 s).
We choose to not time-shift the two Hanford detectors relative to one
another since there may be real correlations due to environmental
disturbances between them.  
The resulting triggers are not correlated at the sites and so
they correspond only to accidental coincidences of noise triggers. 

\subsection{Background result}

A total of 80 time-lags were analyzed to estimate the
background. Specifically the values of the time shifts ranged from
$\Delta t = - 407\; s$ up to $\Delta t = + 407 \;s$ in increments of 
$10 \; s$. The time lags of $\pm 7 s$ were not performed.
We focus here on the signal-to-noise ratio for the triggers from
each detector: $\rho_{\mathrm{L}}$ for L1 and $\rho_{\mathrm{H}}$ for
the Hanford detectors. 

\begin{figure}[ht]
\begin{center}
\includegraphics[height=7cm,width=10cm,clip]{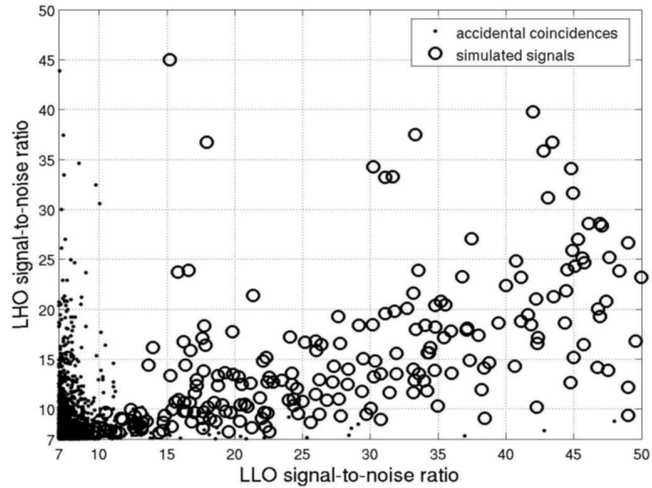}
\end{center}
\caption{\label{f:background}
The accidental coincidences from the time-shifted triggers (dots)
and the triggers from the simulated signal injections (open circles)
are shown.} 
\end{figure}

The distribution
of time-lag triggers in the $(\rho_{\mathrm{L}},\rho_{\mathrm{H}})$ plane
is shown in fig.~(\ref{f:background}). The accidental coincidence 
triggers are plotted as dots in that graph. There is a concentration of those
triggers at the lower left corner of the graph, which corresponds to
the region of low signal-to-noise ratio for all detectors. 
There are also ``tails'' in the distribution, that correspond to triggers 
that have high signal-to-noise ratio in one detector and low 
signal-to-noise ratio in the other. The distribution is quite
different from the equivalent distribution that was observed in the
binary neutron star search \cite{s2bns}, where those
tails of triggers were not present. The presence of these tails 
(and their absence from the corresponding distribution in the binary
neutron star case) can be attributed
to the fact that the $\chi^2$-veto was not applied in this search 
and thus some of the loud instrumental noise bursts that would have been
eliminated by that test have instead survived.

\subsection{Comparison of injections to the background}

It is interesting to compare the accidental coincidence triggers with
triggers that result by recovering simulated inspiral signals added on
the data. The triggers from the simulated signals
are also plotted in fig.~(\ref{f:background})
as open circles.

The distributions of the two types of triggers are quite different.
The injection triggers cluster mainly in the region 
under the diagonal of the $\rho_{\mathrm{H}}$ versus 
$\rho_{\mathrm{L}}$ plane. That is due to the fact that, as was mentioned
in sec.~(\ref{pipeline}), the L1 interferometer was more sensitive than
either of the Hanford interferometers during S2. Consequently, signals
that are detected at both sites have a higher signal-to-noise ratio in
L1 than in H1 or H2. The few injection triggers that are above the 
diagonal correspond to injected signals from binary systems that are
favorably oriented for H1 and H2 but not so for L1, which results to them
having a higher signal-to-noise ratio at the Hanford instruments.

\section{Summary}
\label{summary}

We described the methods used and the preliminary results of the 
first search for binary black hole inspirals that has
been performed in real data. This search, even though similar in
some ways to the binary neutron star inspiral search, 
has some significant differences
and presents unique difficulties. The methods discussed in this 
paper are being used to complete the search in the S2 data and have
given us insight on issues related to the spinning binary black hole
search being performed on the data from the third science run of LIGO.

The fact that the performance and sensitivity of the LIGO interferometers
is improving and the frequency sensitivity band can be extended to lower
frequencies makes us hopeful that the first detection of gravitational
waves from the inspiral phase of binary black hole coalescences may
happen in the near future.
In the absense of a detection, astrophysically interesting results
can be expected by LIGO very soon. It is estimated that at design 
sensitivity the LIGO detectors will be able to detect binary black hole
inspirals in at least 5600 Milky Way Equivalent Galaxies (MWEGs) with
the most optimistic calculations giving 13600 MWEGs
\cite{nutzman}. A science run of 2 years at design sensitivity is
expected to give rate upper limits of less than $10^{-4} \; y^{-1} 
\;MWEG^{-1}$.

\section*{Acknowledgements}

The authors gratefully acknowledge the support of the United States National 
Science Foundation for the construction and operation of the LIGO Laboratory 
and the Particle Physics and Astronomy Research Council of the United Kingdom, 
the Max-Planck-Society and the State of Niedersachsen/Germany for support of 
the construction and operation of the GEO600 detector. The authors also 
gratefully acknowledge the support of the research by these agencies and by the 
Australian Research Council, the Natural Sciences and Engineering Research 
Council of Canada, the Council of Scientific and Industrial Research of India, 
the Department of Science and Technology of India, the Spanish Ministerio de 
Educacion y Ciencia, the John Simon Guggenheim Foundation, the Leverhulme Trust,
 the David and 
Lucile Packard Foundation, the Research Corporation, and the Alfred P. Sloan 
Foundation.


\maketitle
\end{document}